\documentclass[11pt,titlepage]{article}
\usepackage{amsmath}
\usepackage{amsthm}
\usepackage{amssymb}
\usepackage{graphicx}
\usepackage{graphpap,graphics}
\usepackage{fancyhdr}

\usepackage{amscd,amssymb,curves,graphpap,fancyhdr}

\theoremstyle{plain}

\theoremstyle{definition}
\newtheorem{definition}{Definition}[section]
\newtheorem{example}{Example}[section]

\theoremstyle{plain}

\renewcommand{\labelenumi}{(\roman{enumi})}

\newfont{\louisone}{cmssbx10 scaled \magstep3}
\newfont{\louistwo}{cmssbx10 scaled \magstep5}
\renewcommand{\labelenumi}{\arabic{enumi}.}
\begin{document}

\renewcommand{\headrulewidth}{0pt}
\chead{
\hfill {\bf \mbox{Technical Report}}\ \\ \ \\}
\rhead{\thepage}
\addtolength{\headheight}{60pt}
\cfoot{ \louisone  }
\title{\huge A Proposed Quantum Low Density Parity Check Code}

\author{
%\centerline{\NSAseal{100pt}}\\
\ \\
\ \\
Michael S. Postol\\
National Security Agency\\
9800 Savage Road\\
Fort Meade, MD 20755\\
Email: msposto@zombie.ncsc.mil\\
\ \\
\ \\
{\louisone }\\
\date{\today}
}
\maketitle

\bigskip

\section{Introduction}

\hspace{5em} 
Low density parity check codes or LDPC codes were discovered by
Gallager \cite{Gal} in 1962. Because of the lack of sufficient
computing power at the time, these codes were largely ignored until
recently. With the recent discovery of turbo codes and their iterative
decoding techniques, there has been renewed interest in LDPC codes, 
which  can also be decoded iteratively. It turns out that in many cases
LDPC codes perform even better than turbo codes in achieving low bit
error rates for a fixed signal-to-noise ratio.

The problem with LDPC codes is that they are usually very difficult
to encode. In \cite{Kou}, Kou, Lin, and Fossorier address this problem by
constructing LDPC codes based on finite geometries. These codes turn
out to be cyclic codes, so they are very easy to encode.

Since LDPC codes by definition have sparse parity check matrices, we
consider the possibility that quantum CSS codes can be constructed from
them. It is hoped that these sparse matrices will lead to more fault
tolerant decoding techniques, since fewer computations will produce less
error.

In the next section we will define LDPC codes and present a simple
decoding algorithm. Section 3 will define the finite geometries used
in the codes of Kou, Lin, and Fossorier. Section 4 will define these
codes and some generalizations. The final section will give a CSS code
construction based on a helpful suggestion from Shu Lin.

\bigskip

\section{Low Density Parity Check Codes}

\hspace{5em} 
We now define low density parity check (LDPC) codes. 
\medskip
\begin{definition}
An {\it LDPC code} is a binary linear code whose parity check matrix
$H$ has the following properties:
\begin{enumerate}
\item Each row consists of $\rho$ ones.
\item Each column consists of $\gamma$ ones.
\item The number of ones in common between 
any 2 columns, denoted $\lambda$, is no greater than 1.
\item Both $\rho$ and $\gamma$ are small compared to the length of the code.
\end{enumerate}
\end{definition}
\medskip
As an example consider the code with parity check matrix
\[ H=\left( \begin{array}{*{19}{c@{\:}}c}
1 & 1 & 1 & 1 & 0 & 0 & 0 & 0 & 0 & 0 & 0 & 0 & 0 & 0 & 0 & 0 & 0 & 0 & 0 & 0\\
0 & 0 & 0 & 0 & 1 & 1 & 1 & 1 & 0 & 0 & 0 & 0 & 0 & 0 & 0 & 0 & 0 & 0 & 0 & 0\\
0 & 0 & 0 & 0 & 0 & 0 & 0 & 0 & 1 & 1 & 1 & 1 & 0 & 0 & 0 & 0 & 0 & 0 & 0 & 0\\
0 & 0 & 0 & 0 & 0 & 0 & 0 & 0 & 0 & 0 & 0 & 0 & 1 & 1 & 1 & 1 & 0 & 0 & 0 & 0\\
0 & 0 & 0 & 0 & 0 & 0 & 0 & 0 & 0 & 0 & 0 & 0 & 0 & 0 & 0 & 0 & 1 & 1 & 1 & 1\\
\hline
1 & 0 & 0 & 0 & 1 & 0 & 0 & 0 & 1 & 0 & 0 & 0 & 1 & 0 & 0 & 0 & 0 & 0 & 0 & 0\\
0 & 1 & 0 & 0 & 0 & 1 & 0 & 0 & 0 & 1 & 0 & 0 & 0 & 0 & 0 & 0 & 1 & 0 & 0 & 0\\
0 & 0 & 1 & 0 & 0 & 0 & 1 & 0 & 0 & 0 & 0 & 0 & 0 & 1 & 0 & 0 & 0 & 1 & 0 & 0\\
0 & 0 & 0 & 1 & 0 & 0 & 0 & 0 & 0 & 0 & 1 & 0 & 0 & 0 & 1 & 0 & 0 & 0 & 1 & 0\\
0 & 0 & 0 & 0 & 0 & 0 & 0 & 1 & 0 & 0 & 0 & 1 & 0 & 0 & 0 & 1 & 0 & 0 & 0 & 1\\
\hline
1 & 0 & 0 & 0 & 0 & 1 & 0 & 0 & 0 & 0 & 0 & 1 & 0 & 0 & 0 & 0 & 0 & 1 & 0 & 0\\
0 & 1 & 0 & 0 & 0 & 0 & 1 & 0 & 0 & 0 & 1 & 0 & 0 & 0 & 0 & 1 & 0 & 0 & 0 & 0\\
0 & 0 & 1 & 0 & 0 & 0 & 0 & 1 & 0 & 0 & 0 & 0 & 1 & 0 & 0 & 0 & 0 & 0 & 1 & 0\\
0 & 0 & 0 & 1 & 0 & 0 & 0 & 0 & 1 & 0 & 0 & 0 & 0 & 1 & 0 & 0 & 1 & 0 & 0 & 0\\
0 & 0 & 0 & 0 & 1 & 0 & 0 & 0 & 0 & 1 & 0 & 0 & 0 & 0 & 1 & 0 & 0 & 0 & 0 & 1
\end{array} \right) .\]
This example appears in \cite{Gal}. Note that the 2 bottom sections of $H$ are column
permutations of the top section. In this case, we have that the length $n$ of
the code is 20, $\rho=4$, $\gamma=3$, and no two columns have more than 1 
``one'' in common.

Gallager gives 2 iterative decoding algorithms in his paper. The first one is
extremely simple and involves hard decision bit flipping. The second is 
a probabilistic soft decision algorithm. We will describe the first algorithm.
The reader is referred to \cite{Gal} for details.

The bit flipping algorithm is as follows:
\medskip
\begin{enumerate}
\renewcommand{\labelenumi}{Step \ \arabic{enumi}.}
\item Compute the parity-check equations. If they are all satisfied, stop.
\item Find the number of unsatisfied parity-check equations for each bit,
denoted $f_i$ for $i=0, 1, ...,n-1$.
\item Identify the set $S$ of bits for which $f_i$ is above some
predetermined threshold.
\item Flip the bits in $S$.
\item Repeat steps 1 to 4 until all of the parity check equations are 
satisfied or a predetermined maximum number of iterations is reached.
\end{enumerate}
\medskip
\hspace{5em}The problem with LDPC codes is that they are very hard to encode, 
especially when the length $n$ is large. Kou, Lin, and Fossorier solve
this problem by constructing an LDPC code which is actually cyclic. It
is very easy and efficient to encode cyclic codes.

\bigskip

\section{Finite Geometries}

\hspace{5em} We now discuss the finite geometries used in the construction of
Kou, et. al. More information about finite geometries can be found in \cite{MS}.
\medskip
\begin{definition} A {\it finite projective geometry} consists of a finite set
$\Omega$ of {\it points} $p, q, ...$ and a collection of subsets $L, M, ...$ of 
$\Omega$ called {\it lines} satisfying the following axioms. (If $p\in L$, we
say that $p$ {\it lies on} $L$ or $L$ {\it passes through} $p$.)
\begin{enumerate}
\renewcommand{\labelenumi}{\roman{enumi}.}
\item There is a unique line denoted $\left(pq\right)$ passing through any 
2 distinct points p and q.
\item Every line contains at least 3 points.
\item If distinct lines $L$ and $M$ have a common point $p$, and if $q$ and $r$
are points of $L$ not equal to $p$, and  $s$ and $t$ are points of $M$ not
equal to $p$, then the lines $\left(qt\right)$ and $\left(rs\right)$ also have
a point in common.
\item For any point $p$ there are at least two lines not containing $p$, and
for any line $L$ there are at least two points not on $L$.
\end{enumerate}
\end{definition}

\begin{definition} A {\it subspace} of the projective geometry $\Omega$
is a subset $S$
of $\Omega$ such that if $p, q$ are distinct points of $S$, then S contains the
line $\left(pq\right)$. A $hyperplane$ is a maximal proper subspace of $\Omega$.

\end{definition}

We will also use a second type of finite geometry in our constructions.

\begin{definition} A {\it Euclidean geometry} is obtained from a
projective geometry by deleting the points of some fixed hyperplane.
\end{definition}

\begin{definition} A set $T$ of points in a projective or Euclidean geometry
is called $independent$ if for every $x \in T$, $x$ does not belong to the
smallest subspace which contains $T \setminus \{x\}$.
\end{definition}

For example, any 3 points on a line are not independent.

\begin{definition} The {\it dimension} of a subspace $S$ of a projective
geometry is $r-1$, where $r$ is the size of the largest independent set
of points of $S$.
\end{definition}

Our first example of a finite geometry will be the projective geometry
$PG(m,q)$.

\begin{definition} Let $GF(q)$ be a finite field and $m\geq 2$. The points of
$\Omega=PG(m,q)$ are the non-zero $(m+1)$-tuples $(a_0,a_1,...,a_m)$ with
$a_i \in GF(q)$ such that $(a_0,a_1,...,a_m)$ and 
$(\lambda a_0,\lambda a_1,...,\lambda a_m)$ are considered to be the same point
if $\lambda$ is a non-zero element of $GF(q)$. (These are called {\it 
homogeneous coordinates} for the points.)
\end{definition}

$\Omega=PG(m,q)$ is a projective geometry of dimension $m$. There are
$q^{m+1}-1$ nonzero (m+1)-tuples, and each point appears q-1 times, so 
$\Omega$ has \[ \frac{q^{m+1}-1}{q-1}\] points.

The line through the distinct points $a=(a_0,a_1,...,a_m)$ and 
$b=(b_0,b_1,...,b_m)$ consists of the points
\[(\lambda a_0+\mu b_0, \lambda a_1+\mu b_1, ..., \lambda a_m+\mu b_m)\]
where $\lambda$ and $\mu$ are 2 elements of $GF(q)$ which are not both 0.\
A line contains $q+1$ points since there are $q^2-1$ choices for $\lambda$
and $\mu$, and each point has $q-1$ representations.

A hyperplane or subspace of dimension $m-1$ in $PG(m,q)$ consists of
those points $a=(a_0,a_1,...,a_m)$ which satisfy an equation
\[\lambda_0 a_0+\lambda_1 a_1+ ...+ \lambda_m a_m=0\] where the 
$\lambda_i$ are elements of $GF(q)$ which are not all zero. Deleting
the hyperplane \[\{(\lambda_0 a_0, \lambda_1 a_1, ..., \lambda_m a_m)|
\lambda_0 a_0=0\},\] gives points which we can take to be of the form
$(1,a_1,...,a_m)$. We call this set the {\it Euclidean geometry} 
$EG(m,q)$.  $EG(m,q)$ has $q^m$ points which can be labeled as $(a_1,...,a_m)$.

\bigskip

\section{Construction of LDPC Codes From Finite Geometries}

\hspace{5em} We now give the constructions of \cite{Kou} for LDPC codes derived
from the finite geometries of the previous section. Our first example is a
class of LDPC codes based on $EG(2, 2^s)$.

Let $\alpha$ be a primitive element of $GF(2^{2s})$. Each nonzero element of 
$GF(2^{2s})$ can be written in the form $\alpha^i$ for some $i$. We can 
express $\alpha^i$ as a 2-tuple $(b_i, c_i)$, where $b_i$ and $c_i$ are
in $GF(2^s)$ and $\alpha^i=b_i+c_i\alpha$. So $GF(2^{2s})$ can be thought of
as the 2-dimensional Euclidean geometry $EG(2,2^s)$. The point ${\bf 0}=(0,0)$
is called the {\it origin} of $EG(2,2^s)$.

Let $p_0, p_1$ be two linearly independent points in $EG(2,2^s)$. Then the
$2^s$ points of the form $p_0+\beta p_1$ with $\beta \in GF(2^s)$ form a
line passing through $p_0$.

If $p_2$ is linearly independent of both $p_0$ and $p_1$, then the lines 
$\{p_0+\beta p_1\}$ and $\{p_0+\beta p_2\}$ intersect in the point $p_0$.
Any two lines are identical or have no more than one point in common.

Given a point $p_0$ in $EG(2,2^s)$, there are \[\frac{2^{2s}-1}{2^s-1}=2^s+1\]
lines intersecting at $p_0$, including the line $\beta p_0$ passing through the
origin. There are $2^s-1$ lines {\it parallel} to any given line. (I.e. they have
no points in common.) So $EG(2,2^s)$ has $2^s(2^2+1)$ distinct lines.

Given a line $L$ and a primitive element $\alpha$ of $GF(2^{2s})$, let
\[v_L=(v_0,v_1,...,v_{2^{2s}-2})\] be a binary $(2^{2s}-1)$-tuple with $v_i=1$
if $\alpha^i$ is a point on L, and $v_i=0$ otherwise. $v_L$ is called the
{\it incidence vector} of the line $L$.

Now form the parity check matrix $H$ for our LDPC code. H is a 
$(2^{2s}-1)\times (2^{2s}-1)$ matrix whose rows are the incidence vectors
of the \[2^s(2^s+1)-(2^s+1)=2^{2s}-1\] lines in $EG(2,2^s)$ not passing
through the origin.

The parity check matrix H has the following properties:

\begin{enumerate}
\item Each row has $\rho=2^s$ ones since there are $2^s$ points on any
line.
\item Each column corresponds to a non-origin point in $EG(2,2^s)$ and has
$\gamma=2^s$ lines passing through it. (Excluding the line through the
origin.)
\item Any 2 columns have one and only one ``1'' in common (i.e. $\lambda=1$)
since given any 2 points, there is a unique line passing through these 2
points.
\end{enumerate}

\begin{definition} The {\it density} of a matrix H is the ratio r of the total 
number of ones of H to the total number of entries in H.
\end{definition}

The parity check matrix for our code has a density \[r=\frac{2^s}{2^{2s}-1}.\]
The density is very small for large values of $s$.

It turns out that we can construct H by writing down one row and circularly 
shifting to obtain all of the other rows. So we actually have a {\it cyclic}
code, which has an easy encoding algorithm.

We will now give an example. In order to understand it, we will need the
following table of $GF(16)$ generated using the primitive polynomial 
$\alpha^4+\alpha+1=0$.

\begin{tabular}{|c|c|c|}\hline
as a 4-tuple & as a polynomial & as a power of $\alpha$ \\
\hline\hline
1000 & 1 & 1 \\
\hline
0100 & $\alpha$ & $\alpha$ \\
\hline
0010 & $\alpha^2$ & $\alpha^2$ \\
\hline 
0001 & $\alpha^3$ & $\alpha^3$ \\
\hline
1100 & $1+\alpha$ & $\alpha^4$ \\
\hline
0110 & $\alpha+\alpha^2$ & $\alpha^5$ \\
\hline
0011 & $\alpha^2+\alpha^3$ & $\alpha^6$ \\
\hline
1101 & $1+\alpha+\alpha^3$ & $\alpha^7$ \\
\hline
1010 & $1+\alpha^2$ & $\alpha^8$ \\
\hline
0101 & $\alpha+\alpha^3$ & $\alpha^9$ \\
\hline
1110 & $1+\alpha+\alpha^2$ & $\alpha^{10}$ \\
\hline
0111 & $\alpha+\alpha^2+\alpha^3$ & $\alpha^{11}$ \\
\hline
1111 & $1+\alpha+\alpha^2+\alpha^3$ & $\alpha^{12}$ \\
\hline
1011 & $1+\alpha^2+\alpha^3$ & $\alpha^{13}$ \\
\hline
1001 & $1+\alpha^3$ & $\alpha^{14}$ \\
\hline
\end{tabular}

\begin{example}
Let $s=2$. Let $GF(2^{2(2)})=GF(16)$ be generated by the primitive polynomial 
$X^4+X+1=0$. This is the 2-dimensional Euclidean geometry $EG(2,2^2)$ over
$GF(2^2)$

Let $\alpha$ be a primitive element in $GF(2^{2(2)})$ and let $\beta=\alpha^5$.
Then $\{0,1,\beta, \beta^2\}$ form the subfield $GF(2^2)$.

Every line in $EG(2,2^2)$ consists of 4 points. Letting $p_0=\alpha^{14}$ 
gives the line $\{\alpha^{14}+\eta\alpha\}$ where $\eta$ ranges over
$GF(2^2)$. Now the 4 values of $\eta$ are
\[\{0,1,\beta, \beta^2\}=\{0,1,\alpha^5, \alpha^{10}\}.\] The 4 points on the
line are now computed using the table.
\[\alpha^{14}+0\alpha=\alpha^{14}\]
\[\alpha^{14}+1\alpha=\alpha^{14}+\alpha\equiv1001\oplus0100=1101
\equiv\alpha^7\]
\[\alpha^{14}+\alpha^5\alpha=\alpha^{14}+\alpha^6\equiv1001\oplus0011=1010
\equiv\alpha^8\]
\[\alpha^{14}+\alpha^{10}\alpha=\alpha^{14}+\alpha^{11}\equiv1001\oplus0111
=1110\equiv\alpha^{10}\]

So our line is the set \[\{\alpha^7,\alpha^8,\alpha^{10}, \alpha^{14}\}\]
with incidence vector (000000011010001). The parity check matrix for
our LDPC code is now
\[ H=\left( \begin{array}{*{15}{c@{\:}}c}
0 & 0 & 0 & 0 & 0 & 0 & 0 & 1 & 1 & 0 & 1 & 0 & 0 & 0 & 1 \\
0 & 0 & 0 & 0 & 0 & 0 & 1 & 1 & 0 & 1 & 0 & 0 & 0 & 1 & 0 \\
0 & 0 & 0 & 0 & 0 & 1 & 1 & 0 & 1 & 0 & 0 & 0 & 1 & 0 & 0 \\
0 & 0 & 0 & 0 & 1 & 1 & 0 & 1 & 0 & 0 & 0 & 1 & 0 & 0 & 0 \\
0 & 0 & 0 & 1 & 1 & 0 & 1 & 0 & 0 & 0 & 1 & 0 & 0 & 0 & 0 \\
0 & 0 & 1 & 1 & 0 & 1 & 0 & 0 & 0 & 1 & 0 & 0 & 0 & 0 & 0 \\
0 & 1 & 1 & 0 & 1 & 0 & 0 & 0 & 1 & 0 & 0 & 0 & 0 & 0 & 0 \\
1 & 1 & 0 & 1 & 0 & 0 & 0 & 1 & 0 & 0 & 0 & 0 & 0 & 0 & 0 \\
1 & 0 & 1 & 0 & 0 & 0 & 1 & 0 & 0 & 0 & 0 & 0 & 0 & 0 & 1 \\
0 & 1 & 0 & 0 & 0 & 1 & 0 & 0 & 0 & 0 & 0 & 0 & 0 & 1 & 1 \\
1 & 0 & 0 & 0 & 1 & 0 & 0 & 0 & 0 & 0 & 0 & 0 & 1 & 1 & 0 \\
0 & 0 & 0 & 1 & 0 & 0 & 0 & 0 & 0 & 0 & 0 & 1 & 1 & 0 & 1 \\
0 & 0 & 1 & 0 & 0 & 0 & 0 & 0 & 0 & 0 & 1 & 1 & 0 & 1 & 0 \\
0 & 1 & 0 & 0 & 0 & 0 & 0 & 0 & 0 & 1 & 1 & 0 & 1 & 0 & 0 \\
1 & 0 & 0 & 0 & 0 & 0 & 0 & 0 & 1 & 1 & 0 & 1 & 0 & 0 & 0 \\
\end{array} \right) \]

\end{example}
\bigskip 
The code based on $EG(2, 2^s)$ has the following parameters:
\begin{itemize}
\item Length: $n=2^{2s}-1$.
\item Number of parity bits: $n-k=3^s-1$.
\item Dimension: $k=2^{2s}-3^s$.
\item Minimum Distance: $d=2^s+1$, so the code can correct $2^{s-1}$ errors.
\end{itemize}

\begin{example}
Let $s=7$. We have an LDPC code based on $EG(2,2^7)$.

This is a (16383,14197) code with minimum distance 129, so it can
correct 64 errors. 

The parity check matrix $H$ has $\rho=\gamma=128$, $\lambda=1$, and
density r=.007813

The rate $R=k/n$ of this code is $14197/16383\approx.867$ If we increase s,
the rate of the resulting code quickly approaches 1, so these codes are
extremely efficient.

\end{example}

\bigskip
We now describe a family of LDPC codes based on the projective geometries
$PG(2,2^s)$. Let $\alpha$ be a primitive element of $GF(2^{3s})$. Let 
\[n=\frac{2^{3s}-1}{2^s-1}=2^{2s}+2^s+1\]

If $\beta=\alpha^n$, then $\beta$ has order $2^s-1$. The elements 
$0,1,\beta,\beta^2,...,\beta^{2^s-2}$ form the subfield $GF(2^s)$.

Partition the elements of $GF(2^{3s})$ into $n$ disjoint subsets of the
form: \[\{\alpha^i,\beta\alpha^i,\beta^2\alpha^i,...,\beta^{2^s-2}\alpha^i\}
\mbox{\ with\ } 0\leq i\leq n.\]

For each $i$ represent this set as $(\alpha^i)$. For any $\alpha^j \in 
GF(2^{3s})$, if $\alpha^j=\beta^k\alpha^i \mbox{\ with\ } 0\leq i\leq n$,
then we represent $\alpha^j$ by $(\alpha^i)$. The $n$ elements of the form 
$(\alpha^i)$ are taken to be the points of $PG(2,2^s)$. 

If $(\alpha^i)$ and $(\alpha^j)$ are 2 distinct points of $PG(2,2^s)$, then
the line $L$ passing through them consists of points of the form 
$(z_1\alpha^i+z_2\alpha^j)$, where $z_1$ and $z_2$ are elements of $GF(2^s)$
at least one of which is nonzero. Since $(z_1\alpha^i+z_2\alpha^j)$ and
$(\beta^k z_1\alpha^i+\beta^k z_2\alpha^j)$ are the same point, the line $L$
consists of \[\frac{(2^s)^2-1}{2^s-1}=2^s+1 \mbox{ points.}\]

Let $(\alpha^m)$ be a point which is not on the line $(z_1\alpha^i+z_2\alpha^j)$.
Then the lines $(z_1\alpha^i+z_2\alpha^j)$ and $(z_1\alpha^m+z_2\alpha^j)$
intersect at the point $(\alpha^j)$. The number of lines in $PG(2,2^s)$ 
intersecting at the point $(\alpha^j)$ is \[\frac{2^{2s}-1}{2^s-1}=2^s+1\].

There are $2^{2s}+2^s+1$ distinct lines in $PG(2,2^s)$.

Given a line $L$, let $v_L=(v_0,v_1,...,v_{n-1})$ be a binary n-tuple with 
$v_i=1$ if $(\alpha^i)$ is a point of L, and $v_i=0$ otherwise. Then $v_L$
is the incidence vector of the line $L$.

We now can form an LDPC code whose parity check matrix $H$ is a 
$(2^{2s}+2^s+1)\times(2^{2s}+2^s+1)$ matrix whose rows are the incidence 
vectors of the $2^{2s}+2^s+1$ lines in $PG(2,2^s)$.

The parity check matrix H has the following properties:

\begin{enumerate}
\item Each row has $\rho=2^s+1$ ones since there are $2^s+1$ points on any
line.
\item Each column corresponds to a point in $PG(2,2^s)$, and
$\gamma=2^s+1$, since any point has $2^s+1$ lines passing through it.
\item Any 2 columns have one and only one ``1'' in common (i.e. $\lambda=1$)
since given any 2 points, there is a unique line passing through these 2
points.
\end{enumerate}

The density of H is \[r=\frac{2^s+1}{2^{2s}+2^s+1},\]
a very small number for large values of $s$.

As in the case of the code based on $EG(2,2^s)$, we can construct H by 
writing down one row and circularly shifting to obtain all of the other 
rows. So in this case, we also have a cyclic code.

The code based on $PG(2, 2^s)$ has the following parameters:
\begin{itemize}
\item Length: $n=2^{2s}+2^s+1$.
\item Number of parity bits: $n-k=3^s+1$.
\item Dimension: $k=2^{2s}+2^s-3^s$.
\item Minimum Distance: $d=2^s+2$, so the code can correct $2^{s-1}$ errors.
\end{itemize}

\begin{example}
Let $s=7$. We have an LDPC code based on $PG(2,2^7)$. This is a $(16513,14325)$
code with minimum distance 130, so it can correct 64 errors. The parity check
matrix $H$ has $\rho=\gamma=129, \lambda=1,$ and density $r=.007812$. The rate
of this code is $R=.867$, comparable to the code based on $EG(2,2^7)$.
\end{example}

We will now briefly describe some generalizations given in \cite{Kou}.
They involve puncturing, extensions, or higher dimensional geometries.

We first look at puncturing. Consider a LDPC code based on $EG(2,2^s)$.
Choose a line $L$ in $EG(2,2^s)$ and remove the columns of $H$ corresponding
to the $2^s$ points in $L$. We now have a matrix with $2^{2s}-2^s-1$
columns and an all zero row. Removing this row gives a 
$(2^{2s}-2)\times(2^{2s}-2^s-1)$ parity check matrix. 

The new parity check matrix has $\gamma=2^s$ ones in each column. The 
rows have $2^s$ or $2^s-1$ ones depending on whether the lines intersect
the original line $L$. We still have that any 2 columns have exactly one
``1'' in common.

The corresponding code is called an {\it irregular} LDPC code, since not
every row has the same weight.

We can also remove multiple parallel lines to get even shorter codes or 
puncture LDPC codes based on $PG(2,2^s)$.

We can extend our LDPC codes by means of {\it column splitting}. Given
an LDPC code of length $n$ with parity check matrix $H$, we create a new code 
of length $qn$. Our new parity check matrix $H_{ext}$ is formed by replacing
each column of H by q columns with $2\leq q\leq 2^s$ for a code based on 
$EG(2,2^s)$ or $2\leq q\leq 2^s+1$ for a code based on $PG(2,2^s)$.

As an example, for a code based on $EG(2,2^s)$, write $2^s=\gamma q+b$ with
$0\leq b<q$. Each column is the same length as the original and $b$ of them 
contain $\gamma +1$ ones, while $q-b$ of them contain $\gamma$ ones.The ones
are put into the columns in a rotating fashion. We illustrate this technique
with the following example:

\begin{example}
Let $s=2$, so $2^s=4$, and let $q=3$. We then have $4=3\gamma + b$ so 
$\gamma=1$ and $b=1$.

So the column \[\left(
\begin{array}{c}
1\\
0\\
0\\
1\\
1\\
0\\
0\\
1\\
0\\
0\\
1\\
1\\
0\\
0\\
1\\
\end{array}
\right)
\mbox{becomes}
\left(
\begin{array}{*{2}{c@{\:}}c}
1 & 0 & 0\\
0 & 0 & 0\\
0 & 0 & 0\\
0 & 1 & 0\\
0 & 0 & 1\\
0 & 0 & 0\\
0 & 0 & 0\\
1 & 0 & 0\\
0 & 0 & 0\\
0 & 0 & 0\\
0 & 1 & 0\\
0 & 0 & 1\\
0 & 0 & 0\\
0 & 0 & 0\\
1 & 0 & 0\\
\end{array}
\right).
\]
\end{example}

Extension gives a $(2^{2s}-1)\times q(2^{2s}-1)$ parity check matrix $H_{ext}$
with the following properties:
\begin{enumerate}
\item Each row has weight $2^s$.
\item Each column has weight $\lfloor \frac{2^s}{q}\rfloor$ or 
$\lfloor \frac{2^s}{q}\rfloor+1$.
\item Any 2 columns have at most one ``1'' in common.
\end{enumerate}

We can construct extensions of codes based on $PG(2,2^s)$ in a similar manner.
Extending the codes increases the code rate and improves performance. Note
that puncturing and extension can be used in combination.

Finally, we mention that we can generalize all of the codes we have described
for $EG(2,2^s)$ and $PG(2,2^s)$ to codes based on the higher dimensional 
geometries $EG(m,2^s)$ and $PG(m,2^s)$ for $m>2$. See \cite{Kou} for the 
details.

\bigskip

\section{Quantum LDPC Codes}

\hspace{5em} Quantum codes arise in a natural way from classical codes. We
examined the possibility of forming a quantum LDPC code based on the 
finite geometry construction. The quantum construction we used was the
well known {\it Calderbank-Shor-Steane} or {\it CSS} codes. These codes are
described in many places in the quantum computing literature. We will use
the definition found in \cite{NC}.

\begin{definition} Suppose $C_1$ and $C_2$ are $[n,k_1]$ and $[n,k_2]$ 
classical linear codes such that $C_2\subset C_1$ and $C_1$ and $C_2^\bot$
both correct $t$ errors. We will define an $[n,k_1-k_2]$ quantum code 
$CSS(C_1, C_2)$ capable of correcting errors on $t$ qubits. We call this
the {\it CSS code of $C_1$ over $C_2$}. The construction is as follows: If
$x$ is a codeword in $C_1$ define the quantum state $|x+C_2\rangle$ by
\[|x+C_2\rangle\equiv\frac{1}{\sqrt{|C_2|}}\sum_{y\in C_2} |x+y\rangle.\]
Here $|C_2|$ denotes the number of codewords in $C_2$ and $x+y$ denotes 
bitwise addition modulo 2. The quantum code $CSS(C_1, C_2)$ is defined
to be the vector space spanned by the states $|x+C_2\rangle$ for all $x\in C_1$.
Note that $|x+C_2\rangle$=$|z+C_2\rangle$ if and only if $x$ and $z$ lie in
the same coset of $C_2$ in $C_1$. The number of cosets of $C_2$ in $C_1$ is
$|C_1|/|C_2|$, so $CSS(C_1, C_2)$ is an $[n,k_1-k_2]$ quantum code.
\end{definition}

We can think of $C_1$ as correcting the bit flip errors and $C_2^\bot$
as correcting the phase errors. (See \cite{NC} for an explanation of
why this works.) 

In order to construct a quantum version of a LDPC code, we needed to
find families of these codes in which codes of the same length nest
in a natural way, and whose duals are easy to describe. The finite
geometry codes are cyclic, so they have easy to describe duals. The problem
was in finding a way to nest these codes.

In a private conversation, Professor Shu Lin, one of the coauthors of \cite{Kou},
suggested splitting rows of the parity check matrix in a manner similar to
the column splitting extension technique we described in section 4. This 
leads to a code with a larger null space and hence a smaller code. The code
produced is still a cyclic code, so it is still easy to find its dual. 

The technique is best illustrated with an example.

\begin{example}
Consider the LDPC code given in Example 4.1. The code was a $[15,7]$
cyclic code. We gave the parity check matrix $H$ in the example. Now
split the rows of $H$ using $q=2$ to produce a $30\times 15$ matrix
$H_{ext}$. As an example, the first row 
\[ \left( \begin{array}{*{15}{c@{\:}}c}
0 & 0 & 0 & 0 & 0 & 0 & 0 & 1 & 1 & 0 & 1 & 0 & 0 & 0 & 1 \\
\end{array} \right) \]
of $H$ becomes the first 2 rows 
\[ \left( \begin{array}{*{15}{c@{\:}}c}
0 & 0 & 0 & 0 & 0 & 0 & 0 & 1 & 0 & 0 & 1 & 0 & 0 & 0 & 0 \\
0 & 0 & 0 & 0 & 0 & 0 & 0 & 0 & 1 & 0 & 0 & 0 & 0 & 0 & 1 \\
\end{array} \right) \]
of $H_{ext}$. Using Mathematica, we row reduced the resulting matrix and
determined that the new code has a check polynomial $h(x)=x^3+1$ and
that the row space of $H$ has dimension 12, so the new code has dimension
3.

Now let $C_1$ be the original code and $C_2$ be the code generated by 
row splitting. We know that $C_2\subset C_1$. So we have a CSS code where
bit flips are corrected by $C_1$ and phase shifts are corrected by 
$C_2^\bot$. The CSS code is a $[15, 7-3]=[15,4]$ quantum code.

Now $C_1$ is an LDPC code. What can we say about $C_2^\bot$? Since 
$C_2$ has a check polynomial $h(x)=x^3+1$, $C_2^\bot$ has a generator 
polynomial $g^\bot(x)=x^3(\frac{1}{x^3}+1)=x^3+1.$ (See \cite{MS}.) So
the check polynomial for $C_2^\bot$ is \[h^\bot(x)=\frac{x^{15}+1}{x^3+1}=
1+x^3+x^6+x^9+x^{12}\] (The division is taken modulo 2.) We then have
the parity check matrix 

\[ H=\left( \begin{array}{*{15}{c@{\:}}c}
0 & 0 & 1 & 0 & 0 & 1 & 0 & 0 & 1 & 0 & 0 & 1 & 0 & 0 & 1 \\
0 & 1 & 0 & 0 & 1 & 0 & 0 & 1 & 0 & 0 & 1 & 0 & 0 & 1 & 0 \\
1 & 0 & 0 & 1 & 0 & 0 & 1 & 0 & 0 & 1 & 0 & 0 & 1 & 0 & 0 \\
\end{array} \right). \]

This matrix has the following properties:
\begin{enumerate}
\item Each row has $\rho=5$ ones.
\item Each column has $\gamma=1$ ones.
\item Any 2 columns have one and only one ``1'' in common (i.e. $\lambda=1$).
\end{enumerate}

So $C_2^\bot$ is actually an LDPC code. This means that bit flip errors and
phase flip errors are both corrected by LDPC codes.

The density of the parity check matrix for $C_2^\bot$ is $15/45\approx .33$,
which is a little more than the density $4/15\approx .27$ for $C_1$.
\end{example}

\bigskip

\section{Conclusions}

\hspace{5em} We have shown that we can use the finite geometry construction
of LDPC codes to construct CSS codes. The CSS codes fix both bit flip and
phase shift errors with LDPC codes. There is still a need to develop a
general theory describing these codes. It is also hoped that the relatively
simple decoding algorithm for LDPC codes will lead to more fault tolerant
decoding algorithms for these CSS codes. Our initial example shows that
this may be a promising research area.

\end{document}